\documentclass[a4paper, 12pt,draftclsnofoot,onecolumn]{IEEEtran}
\usepackage[T1]{fontenc}
\usepackage{amsmath,amsthm,amsfonts,amssymb,bm,mathrsfs}
\usepackage{geometry}
\usepackage[colorlinks, linkcolor=black, citecolor=black, urlcolor=black]{hyperref}
\usepackage{graphicx,subfigure}
\usepackage{booktabs,multirow,array}
\usepackage{color}
\usepackage{algorithm,algorithmic}

\usepackage{cite}

\begin{document}
\title{Full-Duplex Cellular Networks: It Works!}
\author{Rongpeng~Li, Yan~Chen, Geoffrey Ye Li, Guangyi Liu
	
	\thanks{R. Li and Y. Chen are with Central Research Institute, Huawei Technologies Co., Ltd, Shanghai 201206, China (email: \{lirongpeng, bigbird.chenyan\}@huawei.com).}
	
	\thanks{G. Li is with the School of Electrical and Computer Engineering, Georgia Institute of Technology, Atlanta, GA 30332-0250, USA (email: liye@ece.gatech.edu).}
	
	\thanks{G. Liu is with the Wireless Department, China Mobile Research Institute, Beijing 100053, China (email: liuguangyi@chinamobile.com).}
}
\maketitle

\begin{abstract}
	Full-duplex (FD) communications with bidirectional transmitting and receiving at the same time and frequency radio resource have long been deemed a promising way to boost spectrum efficiency, but hindered by the techniques for self-interference cancellation (SIC). Recent breakthroughs in analog and digital signal processing yield the feasibility of beyond $100$ dB SIC capability and make it possible for FD communications to demonstrate nearly doubled spectrum efficiency for point-to-point links. Now it is time to shift at least partial of our focus to full duplex networking, such as in cellular networks, since it is not straightforward but demanding novel and more complicated interference management techniques. Before putting FD networking into practice, we need to understand that what scenarios FD communications should be applied in under the current technology maturity, how bad the performance will be if we do nothing to deal with the newly introduced interference, and most importantly, how much improvement could be achieved after applying advanced solutions. This article will shed light on these questions.
\end{abstract}

\section{Introduction}
To satisfy the surging traffic demand, mobile networks are facing unprecedented challenges to further improve the efficiency of spectrum usage. Conventionally, mobile networks operate in a half-duplex (HD) mode, which implies only one direction transmission at the same time and frequency radio resource without the extra cost for spatial separation. For example, the base station (BS) can transmit to users (downlink, DL) at one time and frequency radio resource and receive from users (uplink, UL) at another. These time and frequency radio resources are also known as channels. They can be separated by time or frequency dimension, called time-division duplex (TDD) or frequency-division duplex (FDD) mode, respectively.
On the other hand, bidirectional transmission at the same time and frequency resource, or full-duplex (FD) communication has long been dreamed but has been hindered by strong self-interference from a node's transmitter to its receiver, which is as hard as trying to hear a whisper while shouting at the top of your lungs \cite{hong_applications_2014}. In an FD transceiver, the self-interfering signal from its transmitter is usually $100$ dB more stronger than the intended receiving signal. Strong self-interference in an FD system will easily get the radio chain at the receiver saturated \cite{hong_applications_2014} and unable to work properly, not to mention decoding the data.

However, recent breakthroughs in analog and digital signal processing bring positive news to the real application of FD communications. It is now feasible to have up to $110$ dB self-interference cancellation (SIC) capability \cite{bharadia_full_2013}. Therefore, the self-interfering signal is mostly removed with the residual strength reduced to the same level as the signal of interest before going through the decoding chain at the receiver, which makes data decoding feasible. As a result, there have been many real-time FD prototypes reported \cite{bharadia_full_2013,duarte_full-duplex_2010,huawei_full-duplex_2015, chung_prototyping_2015}.

\begin{figure}[htp]
\centering
\includegraphics[width=0.85\textwidth]{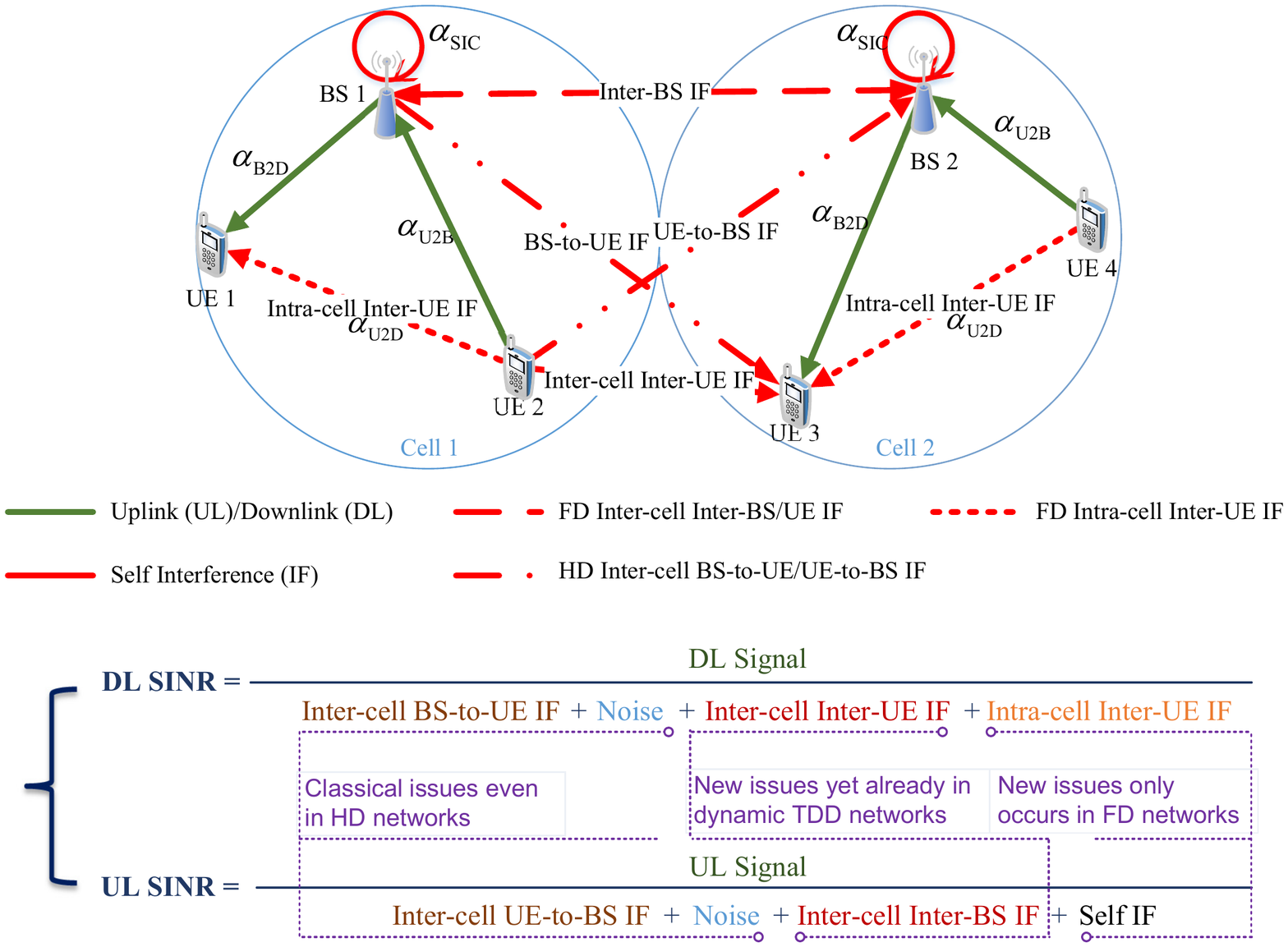}
\caption{An illustration of different types of interference in FD networking.}
\label{fig:interference}
\end{figure}

While roughly doubled throughput has been reported for single-link FD transmission \cite{chung_prototyping_2015}, putting FD communication into networks is not that straightforward. The reasons behind this are two-fold. Firstly, lack of synchronization in transmission direction introduces far more complicated interference than that in both conventional HD networks and dynamic TDD networks\footnote{In dynamic TDD networks, different BSs are designed to have the flexibility to configure different UL and DL subframe patterns, which aims to better adapt to dynamic variations in DL/UL traffic demands on the BS basis.} \cite{shen_dynamic_2012}. Fig. \ref{fig:interference} illustrates different types of interference in FD cellular networks. With only BS working in FD mode, on top of the inter-cell BS-to-UE (user equipment) and UE-to-BS interference that already exists in HD networks, FD networks newly experience inter/intra-cell inter-UE interference, inter-cell inter-BS interference, as well as the residual self-interference after SIC. To make the matter even worse, the interference powers are changing along with user mobility and channel variations. Hence, smart interference management techniques need to be applied to make sure the interference would not eat up the potential benefit from single-link FD communications \cite{duplo_eu_2012, goyal_full_2015, sultan_mode_2015,sabharwal_-band_2014, xin_co-channel_2015,han_full_2014}. Secondly, it is still costly to equip FD functionality with above $100$ dB for all UEs, so most of the UEs may still work in HD mode at least for the near future. Therefore, given the current SIC and interference management capability, it urgently needs careful selection of application scenarios for FD communications, as well as the protocol and algorithm design therein.

\subsection{Scope and Evaluation Framework}
\label{sec:scope}
To apply FD communication to cellular networks, it comes to a wide consensus that macrocell is not a good candidate scenario because of the large transmission power of macro BSs imposed by the large coverage requirement \cite{sabharwal_-band_2014,duplo_eu_2012}. Instead, the architectural progression towards short-range systems, such as small-cell (e.g., picocells) systems where the cell-edge path loss is less than that in macrocell systems, makes the self-interference reduction problem much more manageable \cite{sabharwal_-band_2014}. Simple calculation shows about $140$ dB SIC is required for macrocell scenarios to bring down the transmission signal to a level similar to picocells. Therefore, in this article, we focus on FD cellular networks with pico BSs operating in FD mode while leaving macro BSs and UEs in the HD mode. We consider heterogeneous networks with out-of-band pico BSs (working in different frequency bands) randomly distributed in the coverage of the macro cells. We shall analyze how serious the problem could be if we directly introduce FD communications to pico BSs therein, and how effective different interference management strategies may exhibit.

We select the two important indicators for system performance evaluation, i.e., system spectrum efficiency (SE) and system energy efficiency (EE). The system SE of an FD network is defined as the joint UL and DL total throughput per unit bandwidth. Mathematically, it is given by
\begin{equation}
\mathrm{SE} =\frac{T_{\rm{tot}}^{\rm{UL}} + T_{\rm{tot}}^{\rm{DL}}}{B_{\rm{tot}}},
\end{equation}
where $T_{\rm{tot}}$ and $B_{\rm{tot}}$ indicate the throughput and allocated bandwidth, respectively. Moreover, a UL or DL variable is denoted by the corresponding superscript (i.e., $\rm{UL}$ and $\rm{DL}$). On the other hand, the system EE of an FD network is defined as the joint UL and DL total throughput per joule energy consumed. Here we only consider transmission energy and ignore signal processing energy. Then it could be mathematically formulated as
\begin{equation}\label{eq:EE}
\mathrm{EE} =\frac{T_{\rm{tot}}^{\rm{UL}} + T_{\rm{tot}}^{\rm{DL}}}{E_{\rm{tot}}^{\rm{UL}} + E_{\rm{tot}}^{\rm{DL}}}=\frac{\mathrm{SE}}{P_{\rm{tot}}^{\rm{UL}} + P_{\rm{tot}}^{\rm{DL}}},
\end{equation}
Here $E_{\rm{tot}}$ and $P_{\rm{tot}}$ stands for energy and power consumption, respectively. From Eq. \eqref{eq:EE}, system SE and EE are correlated and since the maximization of total $\mathrm{SE}$ and the minimization of the total $P_{\rm{tot}}$ are usually not achieved at the same time, there exists an interesting SE-EE tradeoff relationship. In this article, we will investigate the behavior of such a relationship in FD networks.

\subsection{Key Findings}
In the rest of the article, we start our evaluation from single-cell FD network. An optimization problem is formulated to maximize the system SE with transmission power and user selection as control variables. We shall show a surprising observation from the analytical solution that for a given pair of UL and DL users, the power control for both the BS and the selected UL UE has binary features, i.e., either transmitting at its full power level or completely muting. Based on this observation, joint power control and user selection problem reduces to a UE pairing one, and interference awareness will play an important role in the UE pairing process. As a step further,  we investigate multi-cell FD networks and identify the dominant interference for different user distribution scenarios based on system level simulator. We will demonstrate through the system SE and EE evaluation that $40$\% to $80$\% SE and EE gains can be achieved with different interference management schemes. In other words, FD networking will work, at least for the considered heterogeneous network setting, with $110$ dB SIC capability.

\section{Single-Cell FD Network}
\label{sec:single}
In single-cell FD network, the interference situation is simple and clear. Compared with traditional HD network, the increased interference is the intra-cell interference from UL UE to DL UE and the self-interference at the transceiver of the BS, as shown in Fig. \ref{fig:interference}. In this case, the problem to maximize the total system throughput of both UL and DL can be formulated as following,
 \begin{equation}
 \label{eq:objective}
 \begin{aligned}
 \max_{u,\ d, \ P_{\rm{BS}}^{d},\ P_{\rm{UE}}^{u}}  \quad f & = \log \left( 1+ \frac{\alpha_{\rm{B2D}}P_{\rm{BS}}^{d}}{N_0+\alpha_{\rm{U2D}}P_{\rm{UE}}^{u}} \right) +\log \left( 1+ \frac{\alpha_{\rm{U2B}}P_{\rm{UE}}^{u}}{N_0+\alpha_{\rm{SIC}}P_{\rm{BS}}^{d}} \right)\\
 s.t. \qquad  &0 < P_{\rm{BS}}^{d} \leq P_{\rm{BS}_{\rm{max}}},\ 0 < P_{\rm{UE}}^{u} \leq P_{\rm{UE}_{\rm{max}}},
 \end{aligned}
 \end{equation}
where $P_{\rm{BS}}^{d}$ and $P_{\rm{UE}}^{u}$ denote the transmission power of the BS (to DL UE $d$) and UL UE $u$, and are limited by the corresponding maximum values $P_{\rm{BS}_{\rm{max}}}$ and $P_{\rm{UE}_{\rm{max}}}$, respectively, $\alpha_{\rm{B2D}}$, $\alpha_{\rm{U2D}}$, and $\alpha_{\rm{U2B}}$ characterize the channels from BS to DL UE, from UL UE to DL UE, and from UL UE to BS, respectively, and will be affected by UL and DL UE pairing. Meanwhile, $\alpha_{\rm{SIC}}$ is determined by the SIC capability of the BS. $N_0$ denotes the noise power.

In this section, we will first dive into the power control problem with only one given pair of UL and DL UEs, and then extend the scope to have multiple pairs and investigate the problem of UE pairing. Finally, we shall give the system level analyses for its SE and EE performance, as well as the tradeoff relationship between them.

\begin{figure}[t]
\centering
\includegraphics[width=0.95\textwidth]{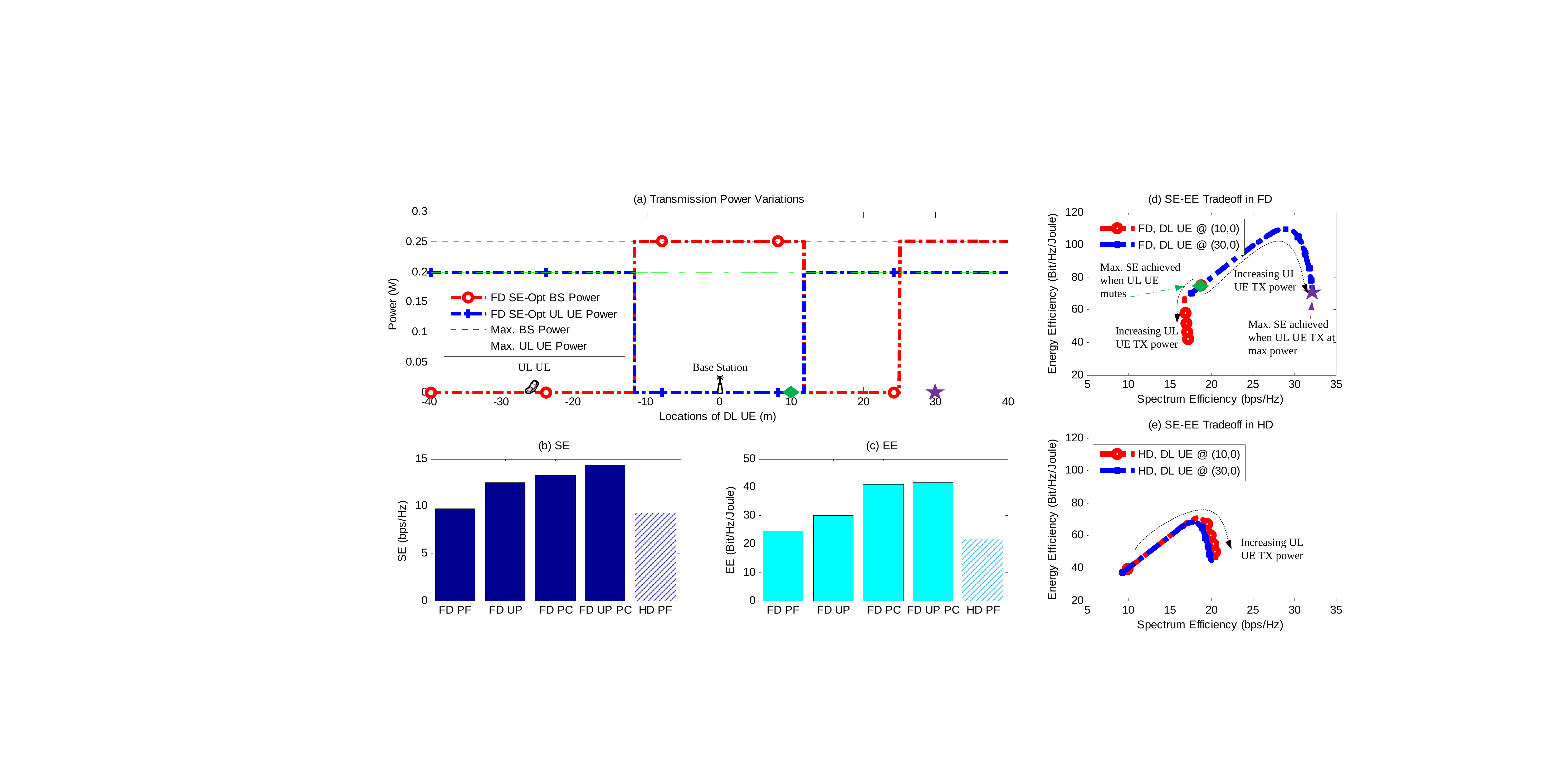}
\caption{(a): The optimal transmission power of BS and UL UE in terms of SE maximization.  (b), (c): The SE and EE performance improvement of Single-Cell FD network over HD network under different interference management strategies. (d): The SE-EE relation in single-cell FD network with a given pair of UL and DL UEs. }
\label{fig:line_seee}
\end{figure}

\subsection{Binary Power Control Solution}
For a given UL and DL UE pair, the problem in Eq. \eqref{eq:objective} reduces to an issue to jointly optimize the transmission power of the BS and the UL UE. Without loss of generality, when the transmission power of UL UE, $P_{\rm{UE}}^{u}$ is fixed, by taking the derivative of Eq. \eqref{eq:objective} with respect to the transmission power of the BS, $P_{\rm{BS}}^{d}$, we can find that there exists at most one minimum value of Eq. \eqref{eq:objective} in the interval $\left[0, P_{\rm{BS}_{\rm{max}}}\right]$ and there is no maximum value inside the interval. Therefore the optimal value of $P_{\rm{BS}}^{d}$ that maximizes the sum rate lies at the two extreme points, i.e., $0$ or $P_{\rm{BS}_{\rm{max}}}$, that is, the BS either transmits no signal or transmits with the maximum power level. Similar result can be obtained for the transmission power of the UL UE. The joint optimization of both, as a generalized case, offers three solution candidates for the pair of $(P_{\rm{BS}}^{d}, P_{\rm{UE}}^{u})$, namely $(0, P_{\rm{UE}_{\rm{max}}})$, $(P_{\rm{BS}_{\rm{max}}}, 0)$, or $(P_{\rm{BS}_{\rm{max}}}, P_{\rm{UE}_{\rm{max}}})$, which shows exactly binary features, i.e., either transmitting at the maximum power or muting. Notably, the first two solutions imply to fall back to HD mode when necessary.

In addition to the analyses above, we also configure a simulation scenario where the BS and UL UE are located at (0,0) and (-25,0), respectively, and set other parameters as in Table \ref{tb:parameter}. By moving DL UE along the horizontal axis from (-40,0) to (40,0), we show the optimal power control solutions for both the UL UE and the BS in Fig. \ref{fig:line_seee}(a). Here, instead of applying our analytical observation above, we perform optimization by exhaustive search. From Fig. \ref{fig:line_seee}(a), to achieve the maximum system SE, the BS and the UL UE either transmit at their maximum power levels or just mute to fall back to HD mode, which is consistent with our analytical observation.

\subsection{Interference-Aware User Scheduling Method}
Given the binary feature of power control, the SE maximization problem in Eq. \eqref{eq:objective} reduces to a UE pairing problem, namely, for given time-frequency resource, how to select one UL UE and one DL UE to properly work together. Basically, there have been many existing scheduling methods in HD network, such as proportional fairness (PF) \cite{goyal_full_2015} and round-robin, which FD network could directly take advantage of. For example, FD network could follow the standard PF procedure to select the DL UE and UL UE independently and pair them. However, the ignorance of inter-UE interference in such a method could degrade the performance. Furthermore, our simulation results will show that interference awareness should be an important feature for the UE pairing process.
There are different levels of interference awareness and also various procedures to achieve that awareness. If we can track the inter-user interference channel fast enough, the short-term interference can be captured, which would be best for performance but with highest overhead in tracking such information. On the other hand, we may only exploit long-term statistics of interference, such as the path-loss, which can be easily derived from the relative user positions. In this case, the interference-aware user scheduling problem turns into a distance-aware one, and has been investigated in \cite{han_full_2014, shao_analysis_2014}.

Here we give an example of a distance-aware joint PF UE pairing algorithm, in which the BS takes turns to select the first user sorted by the PF criteria in UL or DL, and then pair a DL or UL UE with the largest distance. To show the benefit of the binary power control (PC for short) and distance-aware joint PF UE pairing (UP for short) schemes, we simulate a single-cell FD network with $4$ randomly deployed UL or DL UEs. Without loss of generality, the baseline HD network in our simulation is assumed to work in the FDD mode, i.e., UL or DL uses half of the total bandwidth. 
The system SE and EE under different strategies are shown in Fig. \ref{fig:line_seee}(b) and Fig. \ref{fig:line_seee}(c), respectively. From the figures, the FD network shows nearly no gain over the traditional HD network, if we do nothing to control it. However, when either power control or UE pairing is used, the performance gain can be significantly improved, and the joint power control and UE pairing scheme provides around $50$\% and $90$\% boost in system SE and EE, respectively.

\subsection{SE-EE Relationship}
We have discussed the SE and EE performance separately. We will give some insight into their tradeoff relationship in this section. Specifically, we consider one BS with a given UL and DL UE pair but at different locations and find the SE-EE relationship by varying the transmission power of the UL UE from $0$ to $23$ dBm while fixing the transmission power of the BS. Fig. \ref{fig:line_seee}(d) and Fig. \ref{fig:line_seee}(e) demonstrate the SE-EE relationship for FD and HD systems, respectively. From the figures, the shape of SE-EE relation does not change with the UE locations in HD network since there is no interference between UL and DL. However, due to the inter-user interference in FD network, the relative location between UL and DL UEs significantly affects the SE-EE relationship, which confirms the effectiveness of the UE pairing method that we have already seen in the previous section. Moreover, for different UE locations, the UL UE transmits at maximum power or shuts up to maximize the system SE in FD network, which again aligns with our binary power control results in Fig. \ref{fig:line_seee}(a). Finally, for both HD and FD network, the SE-EE relationship is a convex curve, which implies the optimal SE and EE cannot be achieved simultaneously and some tradeoff is necessary. Nevertheless, with advanced interference management strategies, the EE performance can be improved by around $70$\% when the maximized SE is increased by around $50$\%.

\begin{table}[t]
	\centering
	\caption{Main parameters in the system-level simulator, which are compatible with 3GPP TR 36.828 \cite{3gpp_3gpp_2012}.}
	\label{tb:parameter}
	\begin{tabular}{m{5em}m{7em}m{31em}}
		\toprule
		Category  & Sub-Category  & Configuration   \\
		\midrule
		TTI & & 1ms \\
		\hline
		Bandwidth & & Half Duplex: 10MHz; Full Duplex: 20MHz \\
		\hline
		\multirow{4}{*}{Topology} & Macro & 500m-ISD at static positions with 3 sectors \\
		& Pico          & 6 or 12 picos uniformly distribted in 500m-ISD macro's region          \\
		& \multirow{2}{*}{UE}  & \textit{uniform}: 96 or 128 users uniformly distributed in 500m-ISD macro's region \\
		& & \textit{clustered}: 2 or 4 users uniformly distributed in 40m-radius picocell's region\\
		\hline
		\multirow{4}{5em}{Propagation Model} & Pathloss      & Strictly following Table A. 1-3 in 3GPP TR 36.828 \cite{3gpp_3gpp_2012} \\
		& \multirow{2}{*}{Shadowing}     & Macro to Pico: 6 dB; Macro to UE: 10 dB; Pico to UE: 10 dB  \\
		& & UE to UE: 12 dB; Pico to Pico: 6 dB \\
		& Noise Figure  & Macro: 5 dB; Pico: 13 dB; UE: 9 dB \\
		\hline
		\multicolumn{2}{l}{Transmission Power}  & Macro: 46 dBm; Pico: 24 dBm; UE: 23 dBm  \\
		\hline
		\multicolumn{2}{l}{SIC Capability}      & 50 dB to 110 dB, 110 dB by default \\	
		\hline
		\multicolumn{2}{l}{Cell Range Extension (bias) } & 6 dB\\
		\hline
		\multicolumn{2}{l}{Proportional Fairness}  & Window length: 500; Exponent factor: 0.05   \\
		\bottomrule
	\end{tabular}
\end{table}

\section{Multi-Cell FD Networks}
In this section, we investigate multi-cell FD networks. As illustrated in Fig. \ref{fig:interference}, multi-cell FD networks suffer from more complicated interference problems. Therefore, before diving into the design of any specific solutions, we first take a look at how bad the interference situation is and which type of interference is dominant. Then we shall discuss what solution is most effective, especially to deal with the dominant interference, and how much gain we may expect in terms of system SE and EE from FD networks.

Since interference situation is too complicated for multi-cell FD networks, it is impossible to answer the questions above through analyses as in Section \ref{sec:single}. Therefore, we resort to a system-level simulator.

As mentioned in Section \ref{sec:scope}, we consider a non-cochannel heterogeneous networks with macro BSs working in HD mode on one frequency band and pico BSs working in FD mode on another. The deployment of macro BSs and pico BSs is specified in 3GPP TR 36.828 \cite{3gpp_3gpp_2012}. The system parameters are shown as in Table \ref{tb:parameter}. Specifically, seven macro BSs in total are located at vertices or center of a hexagon and pico BSs are randomly scattered in each sector of macro BSs. We consider two ways to drop users, uniform distribution in the coverage of macro BSs, or clustered distribution in the coverage of pico BSs. In the former case, the UE association follows the standard strongest received signal strength (RSS) criterion with $6$ dB cell range bias towards pico BSs. In the latter case, all UEs are served by pico BSs. The results in this article are averaged over $100$ user drops for both cases.

\subsection{Interference Analyses}
\begin{figure}[htp]
	\centering
	\includegraphics[width=\textwidth]{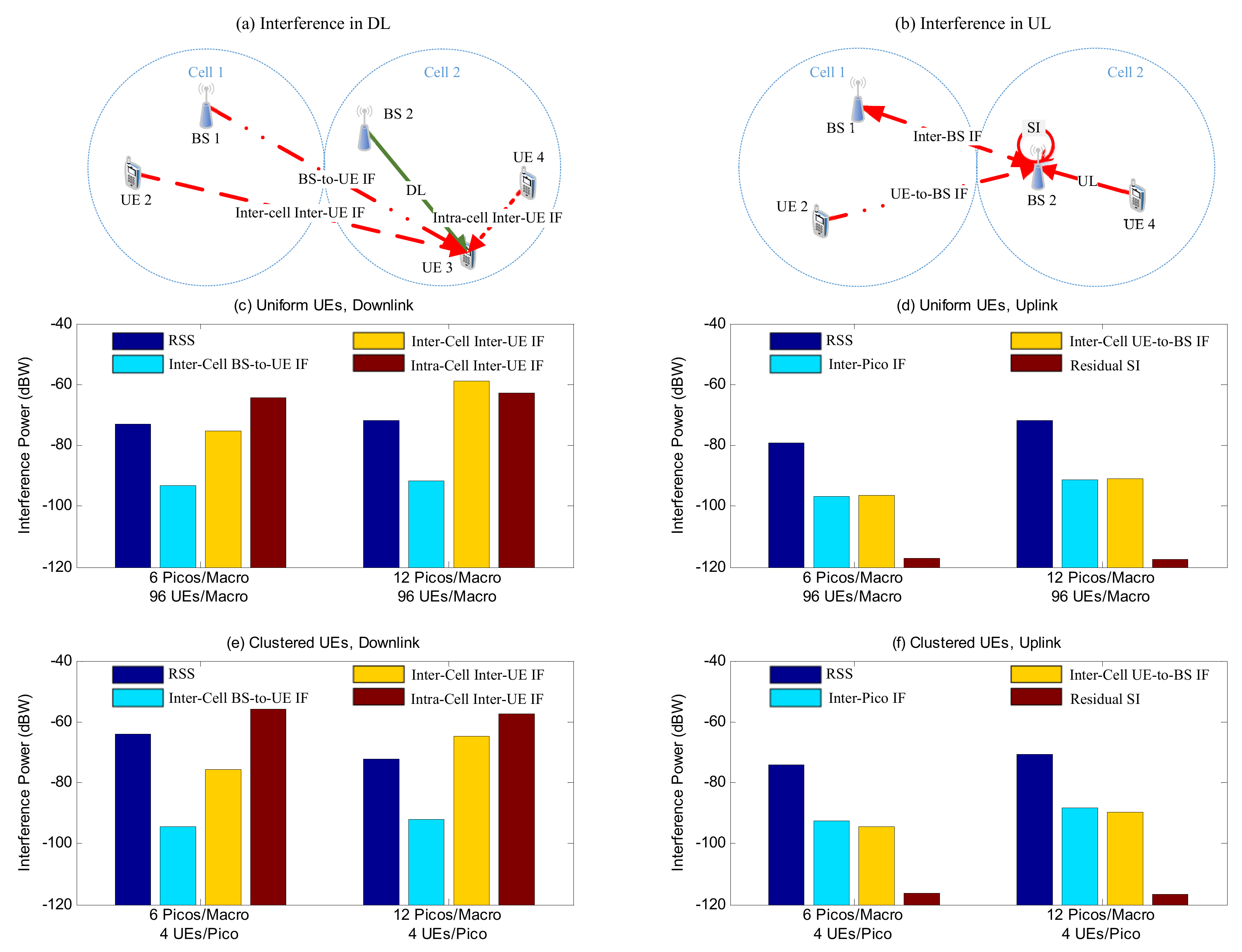}
	\caption{Interference in DL (a) and UL (b) and corresponding powers in two typical scenarios: (c), (d): Uniform UE distribution; (e), (f): Clustered UE distribution. }
	\label{fig:interfResults}
\end{figure}

In this section, we investigate the strength of different types of interference for both uniform and clustered UE distributions by exploiting standard PF scheduling method for UL and DL UEs separately and applying no smart interference management scheme. Furthermore, since the interference situation is different, we present the results separately in Fig. \ref{fig:interfResults}(a) and Fig. \ref{fig:interfResults}(b). We shall see which direction is affected more seriously and which interference is more dominant.

\subsubsection{Uniform UE Distribution}
Fig. \ref{fig:interfResults}(c) and Fig. \ref{fig:interfResults}(d) show the interference powers for the UL and DL UEs, respectively. Two groups of results are shown in each figure, corresponding to two settings of the pico BS density, $6$ and $12$ for each macro sector\footnote{As may be needed for result comparison, when there are 96 uniformly distributed UEs per macro BS, statistically around 2 to 4 UEs are associated to each pico BS.}. 

From Fig. \ref{fig:interfResults}(c), for DL transmission, the strongest interference in most cases is from the UL UE of the same cell, which is a unique problem in FD networks. Meanwhile, DL transmission, on average, is affected more by the inter-cell inter-UE interference than from the inter-pico interference for the first group\footnote{Given the random drop of UEs, the second-order statistic also shows that the inter-cell inter-UE interference has much larger dynamic range that that of inter-pico interference.}. On the other hand, for UL transmission, the interference from neighboring pico BSs is as large as that from UL UEs in neighboring cells, as demonstrated by the first group of results in Fig. \ref{fig:interfResults}(d). Inter-cell interference power increases with the number of pico BSs. Moreover, it also implies that more users in FD networks will incur larger inter-cell inter-UE interference for both UL and DL from Fig. \ref{fig:interfResults}(c) and Fig. \ref{fig:interfResults}(d), which is consistent with our instinction.

\subsubsection{Clustered UE Distribution}
In this scenario, we evaluate the impact of interference when UEs are clustered. The corresponding results are shown in Fig. \ref{fig:interfResults}(e) and Fig. \ref{fig:interfResults}(f). Compared with the uniform case, both intra-cell and inter-cell inter-UE interference in this clustered  case, are significantly larger, as with users being more centralized around pico BSs, the distance between UEs becomes much shorter statistically. Hence, to exploit the potential benefit of FD communications, interference management schemes to cope with inter-UE interference are critical here. Moreover, similar to that in the uniform distribution case, along with the increase in the number of pico BSs, the interference problem becomes more severe.

\begin{figure}[t]
	\centering
	\includegraphics[width=.85\textwidth]{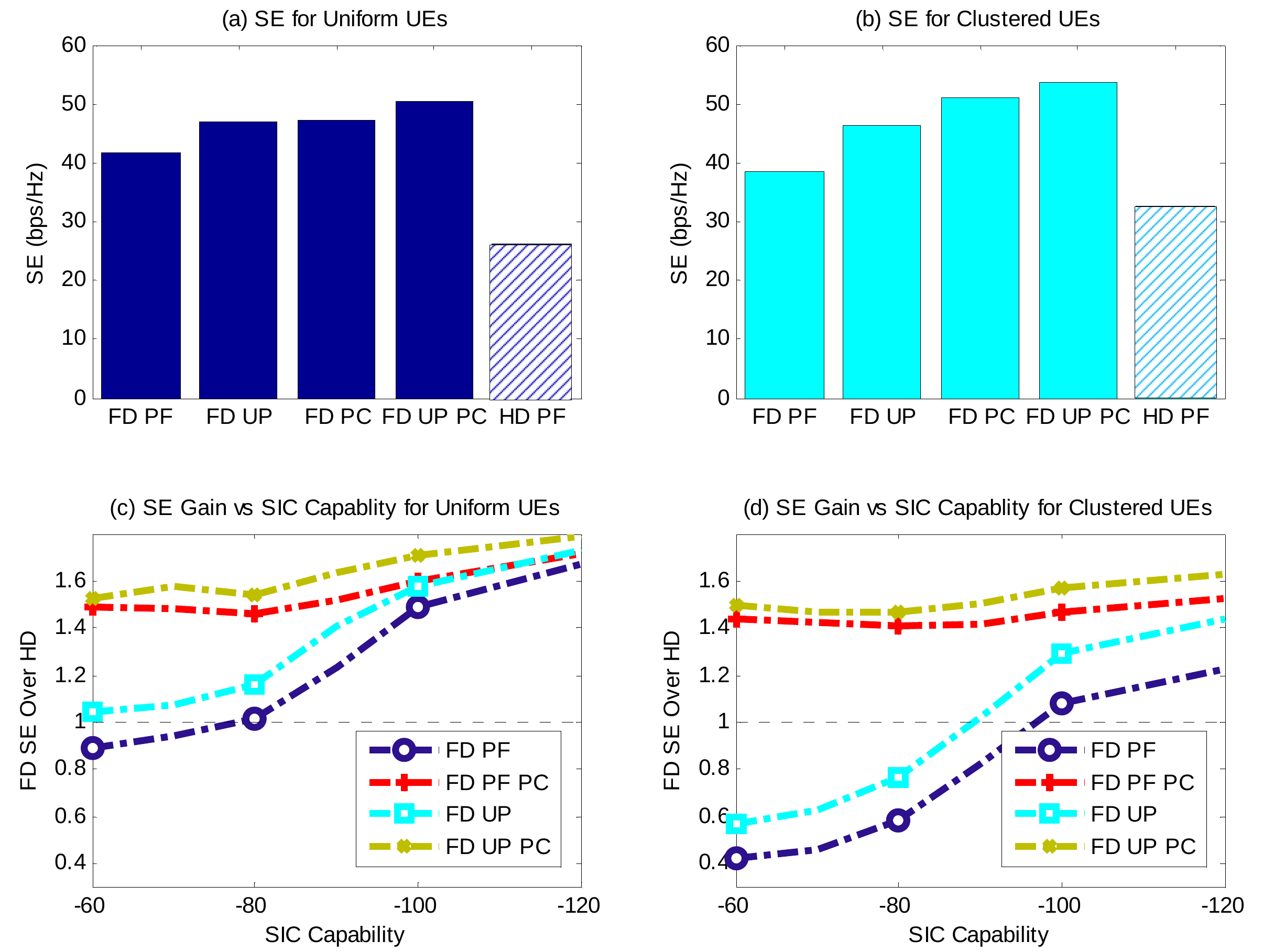}
	\caption{SE performance between uniform UE distribution (6 pico BSs/Macro BS, 96 UEs/Macro BS) and clustered UE distribution (6 pico BSs/Macro BS, 4 UEs/Pico BS). (a), (b): The SE with respect to different interference management schemes, $110$ dB SIC assumed; (c), (d): the SE gain of FD networks over HD networks versus the SIC capability.}
	\label{fig:dlulsic}
\end{figure}


\subsection{Network SE and EE}
In this section, we will investigate how the interference management schemes in single-cell FD network could contribute to improving the multi-cell performance in terms of SE and EE. Specifically, five cased are considered in Fig. \ref{fig:dlulsic} and compared for both UE distribution cases, namely
\begin{itemize}
\item {\it HD PF}: All BSs in HD mode with PF scheduling for UL and DL separately.
\item {\it FD PF}: Pico BSs in FD mode with PF scheduling for UL and DL separately.
\item {\it FD PF PC}: \textit{FD PF} with binary power control (PC) only.
\item {\it FD UP}: \textit{FD PF} with distance-aware joint UL/DL PF UE pairing (UP) only.
\item {\it FD UP PC}: \textit{FD PF} with both PC and UP.
\end{itemize}

From Fig. \ref{fig:dlulsic}(a) and Fig. \ref{fig:dlulsic}(b), under the assumption of $110$ dB SIC capability, for both UE distribution scenarios, positive gains ($58$\% and $18$\% for uniform and clustered case respectively) of FD networking in system SE can be achieved even when no extra interference management strategy is used. This is because in FD case, all bandwidth could be used for both UL and DL UEs, so user diversity improves system throughput. It is encouraging to see extra $20$\%, or $24$\% to $38$\% gain by applying single-cell based power control or UE pairing on top, for uniform case and clustered case, respectively. This verifies the observation earlier that the intra-cell inter-UE interference is most dominant under our setting. From the figure, the gain for the clustered case is higher because the intra-cell inter-UE interference is more severe there as in Fig. \ref{fig:interfResults}. Moreover, extra $35$\% gain can be obtained when these two interference management strategies are combined, which implies that FD networks shall perform the interference-aware UE pairing and even fall back to the HD mode to make sure there is no loss in performance. 

Next, we investigate how the SE gain of FD networks over HD networks with different assumption of the SIC capability. Fig. \ref{fig:dlulsic}(c) shows for the standard PF scheduling method in uniform UE distribution, it needs at least an $80$ dB SIC capability to achieve the sum rate gain of FD networks, and requires a less effective SIC capability if clever interference management schemes are leveraged. Moreover, with the aid of power control method, FD networks could fall back to HD mode whenever necessary to reap a larger SE at some transmission direction (i.e., UL and DL). Hence, it always exhibits performance improvement even when the SIC capability is not so effective. On the other hand, Fig. \ref{fig:dlulsic}(d) demonstrates in clustered UE distribution, stronger SIC capability, around $100$ dB, is needed to mitigate the negative impact of other kinds of interference, such as the intra-cell inter-UE interference.

\subsection{SE-EE Relationship}

\begin{figure}[hbtp]
\centering
\includegraphics[width=0.75\textwidth]{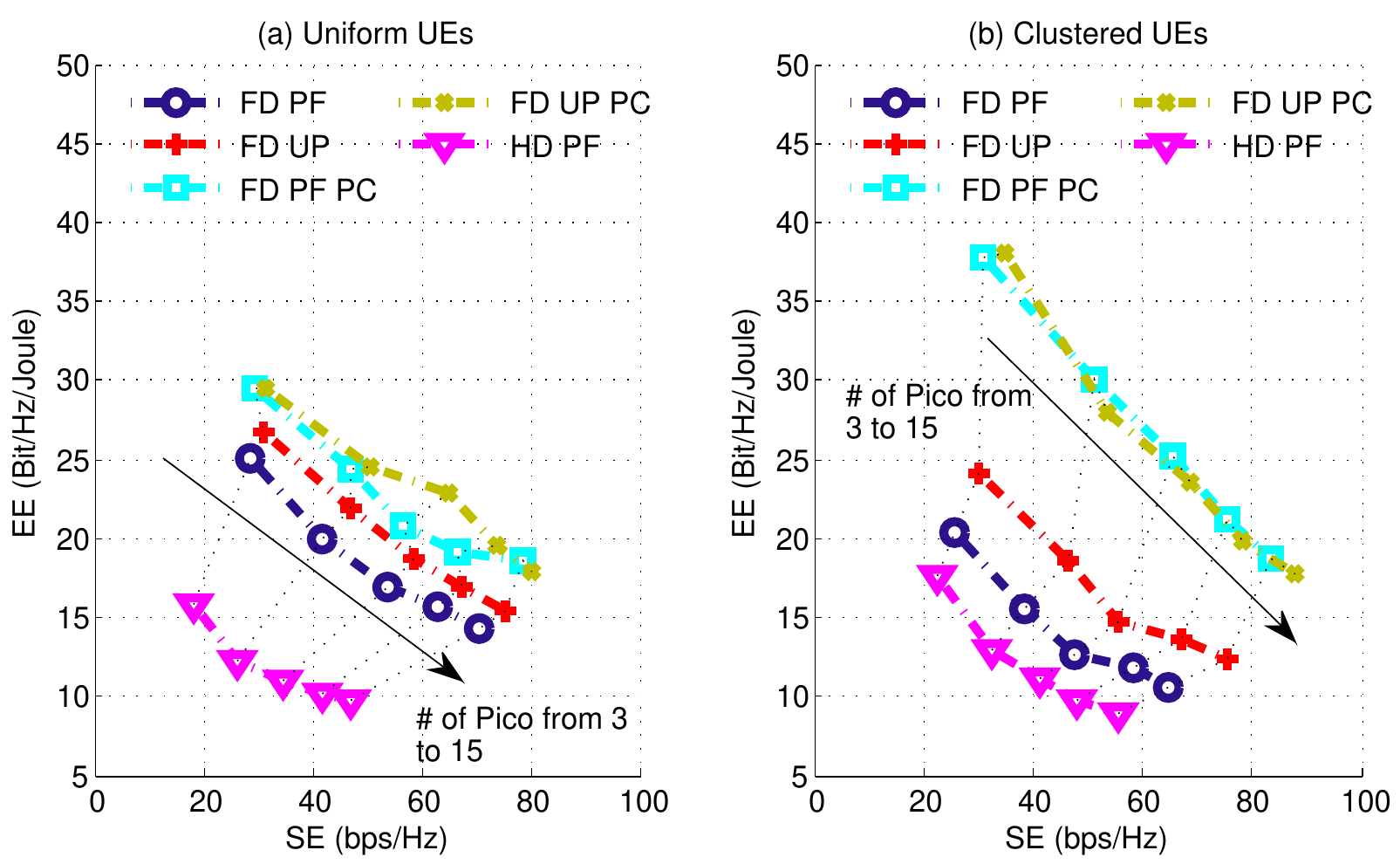}
\caption{The SE and EE performance of FD cellular networks: (a): uniform distribution with 96 UEs per macro BS; (b): clustered distribution with 4 clustered UEs per pico BS.}
\label{fig:seee}
\end{figure}

Fig. \ref{fig:seee} further presents the system SE and EE performance of FD cellular networks. Through the figures, we can observe similar SE-EE tradeoff curves in both UE distribution cases. When the number of pico BSs per macro BS increases from 3 to 15, it leads to distinct variation trends in SE and EE. It is because more pico BSs imply a higher frequency reuse ratio, thus leading to a larger SE. However, deploying more pico BSs also add to the total power consumption and incur larger inter-cell interference. Consequently, as the SE gain cannot compensate the loss in interference and power consumption, the EE decreases. Meanwhile, in addition to the benefit to SE performance improvement already validated in Fig. \ref{fig:dlulsic}, Fig. \ref{fig:seee} implies FD networks could benefit system EE as well. For both UE distribution cases, FD networks with standard PF scheduling method could bring $20$\% larger EE than HD networks. By exploiting power control and UE pairing methods, the EE performance improvement could be as large as $120$\%.

\section{Conclusions and Future Works}
In this article, we have suggested equipping pico BSs with FD functionality will be most practical and promising for FD networking in cellular networks. Starting with a single-cell FD network, we have highlighted the binary power control feature for any given UL and DL UE pair and reduced the system SE optimization problem to a UE pairing problem, in which we have demonstrated the importance of interference-awareness. Based on our interference analysis results, intra-cell inter-UE interference will be most dominant under our setting. Therefore, we have further combined the UE pairing scheme based on distance-aware joint PF scheduling and the binary power control scheme as potential interference management solution for multi-cell FD networks. The system-level simulation has proven $80$\% and $50$\% SE gains over the traditional HD networks for the uniform and clustered cases, respectively, under $110$ dB SIC capability assumption. From here, we can conclude, FD works for cellular networks!

However, there still exist demanding challenges to solve ahead, including the combination of FD functionality with multiple-input mutiple-output (MIMO) system design, the protocol and algorithm design to take advantage of the interference cancellation at the receiver or even to combine with the non-orthogonal multiple access schemes \cite{lu_prototype_2015}, the extension to UEs with FD capability in both cellular and device-to-device (D2D) communications.

\section*{Acknowledgment}
We want to express our sincere thanks to Eddy Hum, Huan Wu, and Moshiur Rahman from Huawei Canada for their  insightful comments to improve the quality of this article.
\bibliography{IEEEabrv,draft}
\bibliographystyle{IEEEtran}
\end{document}